# Cybersickness Assessment Framework: Towards a Standardization of Experiments

Nana Tian[1,2*†], Elif Kurtay[1†], Dylan Vairoli[1†], Adriano Viegas Milani [1†], Ronan Boulic[1,2]

[1*]School of Computer and Communication Science, École Polytechnique Fédérale de Lausanne, Lausanne, Vaud, Swizterland.
[2*]Immersive Interaction Group, École Polytechnique Fédérale de Lausanne, Lausanne, Vaud, Swizterland.

*Corresponding author(s). E-mail(s): nana.tian@epfl.ch;
Contributing authors: elif.kurtay@epfl.ch; dylan.vairoli@epfl.ch; adriano.viegasmilani@epfl.ch; ronan.boulic@epfl.ch;
†These authors contributed equally to this work.

**Abstract**

Investigating cybersickness (CS) in virtual reality (VR) often requires significant resources to create the VR environment and manage other experiment-related aspects. Additionally, slight differences in VR content across studies can lead to conflicting results. To address these challenges, we pro- pose a standardized assessment framework to facilitate cybersickness research. The main goal is to enable consistent and comparable CS-related experiments. By establishing this common foundation, researchers can better evaluate and compare the impact of various factors on cybersickness. We pro- vide a comprehensive explanation of the conceptual designs, detail the technical implementation, and offer instructions for using the proposed framework. Lastly, we conclude by discussing the limitations and potential avenues for future development.

Keywords: Virtual Reality, Cybersickness, Testbed, Framework

## 1 Introduction

Cybersickness, also widely recognized as visually induced motion sickness, has emerged as a significant obstacle impeding the widespread adoption of virtual reality (VR) technologies. The adverse effects of cybersickness, such as nausea, dizziness, and general discomfort, not only detract from the immersive VR experience but also limit the duration of exposure for users. [1].

### 1.1 Conflicting results

The field of cybersickness research faces a persistent challenge characterized by conflicting find- ings across numerous studies. Despite a con- siderable volume of published papers investigat- ing factors influencing cybersickness, reduction techniques, and predictive models, the incon- sistent outcomes complicate the prospects for future research, hindering the acquisition of clear insights and guidance [1]. For example, when examining the influence of gender on cybersickness, research studies have produced divergent findings. Some studies, suggest that females are more susceptible to cybersickness than males [2][3][4][5]. Conversely, other studies, indicate that there is no significant difference in susceptibil- ity between genders [2][3][4][5][6]. Similar conflicts have emerged regarding the influence of factors like age, past experiences, and controllability on cybersickness[1]. In our investigation, we also found diverse outcomes in studies employing iden- tical physiological measures. For instance, one study by Garcia et al. reported that individuals experiencing more severe cybersickness exhibited lower heart rates [7]. Conversely, Gavgani's study

[8] found an increase in heart rate among partic- ipants who experienced cybersickness[9]. Interestingly, Wibirama's paper observed both an increase and a decrease in heart rate among different sub- jects [9]. In contrast, a study by Guna did not find a significant correlation between heart rate and cybersickness [10]. Similarly, research papers that focus on evaluating cybersickness reduction techniques, such as field of view (FOV) reduction and teleportation, also present inconsistencies in the reported results [11][12][13][14]. The incon- sistencies observed in the results of cybersickness research can be attributed to three key perspec- tives: experimental settings, human factors, and hardware settings. Firstly, variations in experi- mental settings, such as the utilization of different virtual reality (VR) stimuli, can lead to diverse levels of cybersickness. The content presented in VR experiences has a substantial impact on the degree of induced cybersickness. However, due to the lack of an established framework or compre- hensive VR game database, researchers often find themselves having to develop their own VR games or select from existing ones, introducing variabil- ity among studies. Secondly, individual differences among participants selected for the studies sig- nificantly influence the final outcomes due to variations in individual susceptibility. Factors like gender, past experiences, and history of motion sickness could play a crucial role in determin- ing the level of cybersickness experienced. How- ever, previous research papers either lack detailed reporting of these individual factors or fail to adequately control them, adding another layer of complexity to the interpretation or comparison of results. Lastly, discrepancies in hardware settings, such as the types of VR headsets used, the quality and positioning of physiological sensors, can have an impact on the research outcomes.

## 1.2 Lack of transparency

A significant challenge encountered in previous research studies is the inadequate provision of details and transparency regarding the VR appli- cations utilized and the collected data. In many papers, only videos or screenshots of the VR games are provided, with very few offering open access to their customized games, with the major- ity relying on commercially available VR games[1]. This lack of transparency greatly hampers the reproducibility of research findings and further impedes the synthesis of results across multiple studies. As an illustration, a recent study posits that the lack of gender difference in cybersick- ness may be attributed to the VR content not being potent enough to elicit a substantial degree of cybersickness [6]. Recognizing the importance of transparency in scientific research, it is essen- tial to address this limitation and promote open access to VR applications.

Hence, addressing these issues by establishing standardized frameworks for VR stimuli, ensuring detailed reporting and control of human factors, and maintaining consistent hardware settings is central in advancing our understanding of cyber- sickness and facilitating more reliable and com- parable research outcomes. Given that researchers have exclusive control over participant selection and hardware applications, our primary objec- tive is to undertake the design of a standardized framework. This framework aims to mitigate the significant variations in VR content, facilitate improved comparisons among research outcomes, and more importantly, reduce prolonged develop- ment cycles. By doing so, we can enhance the consistency and reliability of the results, enabling a more comprehensive understanding of cybersick- ness and its associated factors.

## 1.3 Targeted software design principles

With these goals in mind, we recall the key soft- ware design principles and desirable features that guided the definition of our framework:

- **Abstraction:** hiding unnecessary implementa- tion details and exposing only relevant inter- faces. This allows users of the framework to interact with it at a higher level, simplifying the usage and reducing potential complexities.
- **Modularity and separation of concerns:** promoting reusability and maintainability. Each component should have a well-defined respon- sibility, making it easier to add or update functionalities without affecting other parts of the system.
- **Extensibility:** accommodating future require- ments or changes. Design with flexibility in mind, allowing developers to add new features or functionalities without major rewrites or disruptions.
- **Advanced Experiment Control:** for design- ing sophisticated and comprehensive manage- ment of experimental conditions, parameters, and variables in scientific research or testing environments.
- **Minimal-Coding:** for facilitating fast and easy creation of new functionalities.

## 1.4 Objectives

The experiment setup involves a user interface (UI) control panel for experiment settings with embedded Fast motion sickness questionnaire in voice (FMS) and data logging system for fur- ther data analysis. The framework's default capa- bilities include two basic games with different themes, options for controlling locomotion within virtual environments, techniques to reduce visual motion sickness (Vision CS reduction techniques), and flexibility in testing different scene arrange- ments. It is adaptable to different versions of the Unity game engine and supports various current head-mounted display (HMD) devices, allowing for versatility in research setups. Furthermore, the framework is designed to be extensible, with the ability to incorporate new features, priori- tize user-friendliness, and maintain readable code for easy maintenance and future enhancements. The subsequent sections provide a comprehensive explanation of the details.

## 2 Related work

### 2.1 Test frameworks in VR studies

Numerous projects have emerged with the aim of streamlining various aspects of VR research ; we summarize their key features in Table 1. To address the challenges associated with envi- ronment creation, one such project is VREX, developed by Vasser et al. [15]. VREX is a Unity project that incorporates an intuitive menu sys- tem for creating indoor environments, specifically catering to furnished rooms for psychological experiments. Another tool in this realm is the Locomotion Usability Test Environment for Vir- tual Reality (LUTE) by Sarupuri et al. [16]. LUTE provides a codeless environment creation tool that simplifies the time-consuming task of generating diverse environments for testing loco- motion techniques. However, the environments offered by LUTE mainly consist of straight roads with occasional turns, which may not be ade- quate for cybersickness research. Regarding tools related to experiment design, Bebko et al. intro- duced bmlTUX [17], a Unity Package that focuses on facilitating the technical aspects of experi- ment organization and evaluation. It provides functionalities for handling variables, randomiza- tion, counterbalancing, and more, along with a preset system 7 for different experiment designs. Another toolkit is Ubiq-exp by Steed et al. [18], which builds upon the Ubiq library and simplifies experiment creation with a particular emphasis on multiplayer aspects, including avatars, voice communication, and remote experiments. Both tools are designed as packages to be integrated into existing projects and do not encompass a comprehensive bundle comprising environments, techniques, or cybersickness-related experiments. GingerVR, proposed by Samuel Ang and John Quarles [19], is a repository of popular visual cybersickness reduction techniques derived from existing literature. These techniques have been summarized and implemented in Unity, elim- inating the need for researchers to repeatedly develop commonly used techniques from scratch and addressing inconsistencies between cus- tomized implementations found in the literature. In the realm of locomotion technique evaluation, Cannavò et al. introduced the Locomotion Eval- uation Testbed VR (LET-VR) [20]. LET-VR offers a range of locomotion techniques and corresponding experiments, employing a protocol and scoring system to rank them based on specific requirements. While LET-VR primarily focuses on locomotion techniques, it lacks the explicit emphasis on cybersickness found in GingerVR and our proposed Cybersickness Assessment Framework (CSAF).

## 2.2 Evaluation of CSAF V1.0

An initial iteration of the Cybersickness Assessment Framework (CSAF 1.0), from Milani et al.[22], addressed some of our objectives but still exhibited a few limitations that we briefly review now. For the convenience of the readers, the definitions of the main technical terms appear in Appendix A.

### 2.2.1 Genericity

- **Adaptability to new scripts**
  The framework allowed for seamless integration of any script, provided that the script's fields can be serialized by the saving system. This capability enables the direct creation and saving of presets for the newly integrated script. Furthermore, researchers have the option to share the newly created features and their corresponding presets with others, facilitating collaboration and knowledge exchange within the research community.
- **Structuration with hierarchy of presets**
  From a technical perspective, it was feasible to generate presets for an entire category or the entire scene simultaneously, rather than limiting it to a specific script. However, in practical terms, this feature had limited usefulness due to its requirement for the GameObjects to possess identical names as those specified in the saved presets. Consequently, any modifications made to the object's name would render the preset obsolete and ineffective.
- **Shareable presets**
  Given that presets were stored as JSON files, it was conceivable to share them with third parties, enabling them to directly apply the preset. However, this approach did not uniformly function in all scenarios. While it held true for individual script presets, the same principle did not apply to scene presets. As scene presets relied on specific names, the recipient would need to possess an identical scene hierarchy in order to utilize the shared preset effectively. If any modifications were made to the scene structure by either the recipient or the original owner, the preset would become obsolete and outdated, rendering it impractical for use.

### 2.2.2 Limitations

- **Maintenance**
  As recalled in the preceding section, the framework's dependency on object names weakened the presets genericity. The inability to rename objects within the scene posed a significant constraint for framework users. Additionally, the process of incorporating a new category into the framework proved to be a complex undertaking, necessitating the creation of a new intricate Manager and the corresponding definition of its methods.
- **Usability**
  It was not feasible to visually inspect the contents of a preset without running the application. Additionally, modifying the values within the preset was challenging due to their storage in a JSON file, which lacked human-readability. This difficulty was particularly pronounced when dealing with non-primitive types, as successful modification without errors is seldom achieved. Moreover, there was no means to determine the specific script or category associated with a preset. For instance, if a preset was named *default* it was impossible to discern which script it served as the default preset for, unless redundancy was introduced by naming it *default your script* Lastly, the framework lacked the capability to generate a list of available presets for a given script.

In summary, while the overarching objectives of CSAF 1.0 was aligned with the goals of this paper, its implementation fell short in terms of maintainability and usability, making it unsuitable for easy adaptation by other users for their own work. Consequently, a significant portion of the effort presented in this paper is dedicated to a comprehensive overhaul of the framework's foundation, with the aim of establishing a usable, functional, and reliable system (referred to as CSAF 2.0 in the rest of the paper).

Table 1: Comparison table measuring the different versions of the CSAF with previous studies[21]

| | VREX[15] | bmlTUX[17] | Ubiq-exp[18] | GingerVR[19] | LUTE[16] | LET-VR [20] | CSAF 1.0[21] | CSAF 2.0 |
|---|---|---|---|---|---|---|---|---|
| Environment Customization | ✓ | | | | ✓ | | ✓ | ✓ |
| Multiple environments | ✓ | | | | ✓ | ✓ | > | ✓ |
| Environment Creation | ✓ | | | | | | > | ✓ |
| CS Reduction Techniques | | | | ✓ | | | ✓ | ✓ |
| Locomotion Techniques | ✓ | | | | | ✓ | ✓ | ✓ |
| GUI | ✓ | ✓ | | | ✓ | | ✓ | ✓ |
| Minimal-Coding | ✓ | ✓ | | | ✓ | ✓ | ✓ | ✓ |
| Tutorials Preset | ✓ | ✓ | ✓ | | | ✓ | > | ✓ |
| System | | ✓ | | | | | ✓ | ✓ |
| Designed for Expansion | | ✓ | ✓ | | | | ✓ | ✓ |
| CS related Experiment | | | | | | ✓ | ✓ | ✓ |
| Advanced Experiment Control | | ✓ | | | | ✓ | > | ✓ |
| Multiplayer Experiment | | | ✓ | | | | X | X |
| Logging | ✓ | ✓ | ✓ | | | ✓ | ✓ | |
| Grading system | | | | | | ✓ | X | X |
| Individual susceptibility | | | | | | | ✓ | ✓ |
| Tool Format | Project | Package | Package | Repository | Repository | Project | Project | Project |

## 3 General structure of CSAF 2.0

Figure 1 summarizes the framework structure. We also list a series of terms to assist the understand- ing of the technical backgrounds.

- **Extension**: In this framework, an extension is an abstract class that adds new functionalities to an existing "Monobehaviour" by being added as a component to a "GameObject." It is a generic type that takes the "Monobehaviour" it extends as a type parameter, allowing it to be bound to a specific "Monobehaviour" and dis- play all available extensions for GameObjects in the scene.
- **Preset**: In this framework, a "Preset" is directly associated with each extension and is given as a type parameter, primarily to cre- ate a custom editor. The purpose of the preset is to encompass all parameters that one may wish to share or save for the related exten- sion. These "Presets" are essentially "Script- ableObjects" that are specialized with a type parameter matching their respective extension, allowing for simple creation of diverse variants of presets. It is worth mentioning that this term is an customized version of the general term "Presets" that is defined in the Appendix.
- **Presettable**: In this framework, "Presettable" refers to objects that have the capability to be defined as "Presets."
- **ScriptableObjects**: "ScriptableObjects" in Unity are data containers that allow developers to create custom, reusable, and serializable data objects that can be easily shared and modified in the Unity Editor.

### 3.1 Setup window Conceptual Design

In the subsequent subsections, we present a com- prehensive conceptual design of the setup window, which includes the four major categories of CSAF 2.0, namely the Experiment, Environment, Vision and Locomotion. Apart from these, we also pro- vide two demo scenes for easy initialization of the framework.

The categories which presettables fall under can also be referred to as an option since these functionalities form options the user can add, remove, or change for their experiment.

The framework's default options are shown in the setup window by their names, whether they are included in the current scene or not. If the scene does not have the extension object of an option, then it can be created and later modified from the same window. In addition, it is possible to create new extensions or presettables to add new options to the experiment, which is explained in detail in the following subsections.

#### 3.1.1 Scenes

Typical locomotion in VR applications can be categorized into two distinct types (Passive and Active). To facilitate an accessible entry point for researchers into this framework, we have incor- porated two fundamental scenes in addition to the traditional interactive elements found in VR games. These scenes can be readily accessed by searching for their respective names ("Coin" or "Shooting") within the Unity project's Assets hierarchy.

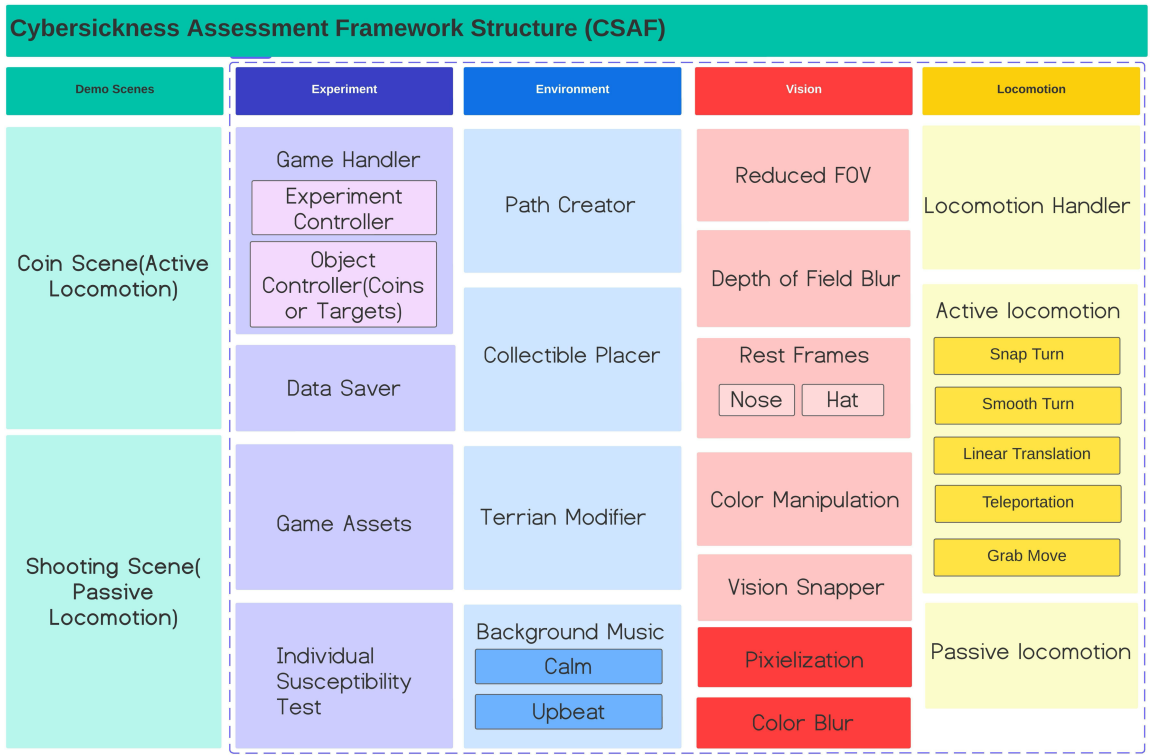

**Fig. 1**: CSAF 2.0 key structures. Two demonstration scenes are included to facilitate an easy initialization of the framework. Each scene is organized into four key categories, encompassing the principal functionalities of the framework.

The "Coin" scene offers an active locomotion approach, akin to the classic Mario game, where users employ controllers to navigate and collect coins dispersed throughout the environment. The coin generation is managed through a preset script within the framework, resulting in a randomized distribution of coins. On the other hand, the "Shooting" scene embraces a passive locomotion style, inspired by the renowned VR game "Pistol Whip." In this setting, users do not exert direct control over their movement, experiencing pre- defined motion paths while engaging in shooting tasks.

### 3.1.2 Experiment

The experiment module comprises essential features related to the defi- nition of an experiment, such as the Game Handler, Data Saver, Game Assets, and the Individual Susceptibility Test. The game handler encompasses the experiment con- troller and object controller, allowing researchers to customize aspects like the number of coins, experiment duration, and Fast Motion Sickness Scale (FMS) parameters through presets[23]. The data saver automatically records user information like head rotation and position in a CSV file for future analysis. The game assets include audio clips that can be played when a coin is collected. Finally, the Individual susceptibility test consists of a brief evaluation to measure an individual's susceptibility to specific factors, like rotation in different axes and linear acceleration along vari- ous directions. Additionally, there are four scenes

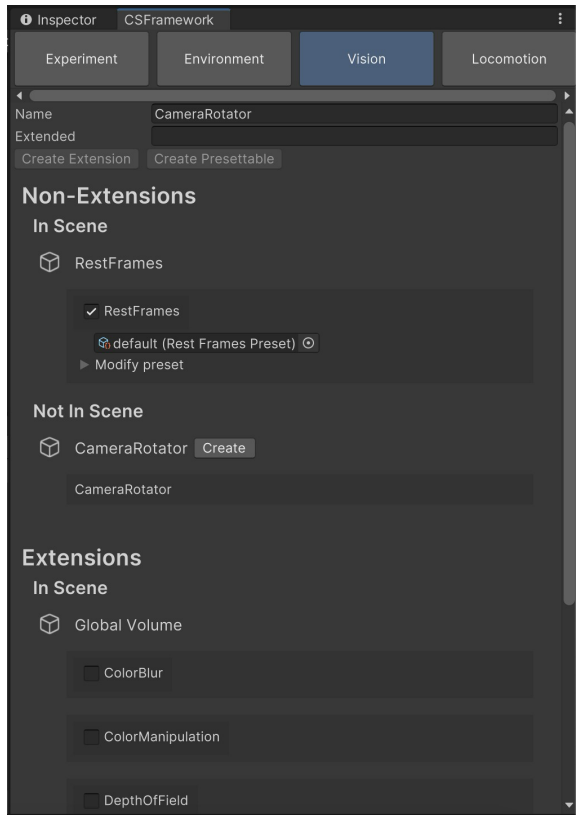

**Fig. 2**: CSAF 2.0 setup window in Unity

that could be selected based on their design and complexity, see Figure 5

**Individual susceptibility test I** There is also the PersonalSensitivityTest option avail- able for each scene. This test is performed by disabling participant controllers in the game and exposing the participant to linear and rotational passive movement in each three axes and 3 differ- ent orders. An indicator alerting the participant of which turn axis will be the next one is shown. After the three combinations of the axis turn- ing completes, the test returns to the main menu screen of the game. The repetition of the sin- gle axis turns, the time of each single axis turn, the time waited after each axis turn, and the time waited between each three axis turns can be modified. The rotational test moves the Camera 360 degrees in each of the three rotational axes, *namely pitch, roll, and yaw*. The linear translation is performed in lateral, vertical and longitudinal axes.

Furthermore, just for the Individual Sensitiv- ity Test, a separate scene (with a space theme) is created. The reason for creating a new envi- ronment is due to a more equalized input of the field of view while rotating in each axis. However, this option is not removed from other scenes so that the freedom of using the personal sensitivity test can be easily added to other environments or scenes prepared by users.

The reason this small test is included inside the framework is that personal susceptibility to VR has been suggested to be an important indi- cator of symptoms in participants[1]. We wanted to include a test on movement susceptibility, since it is shown in the literature that the optic flow direction has a significant influence on the symp- toms elicited during cybersickness[24]. The effects of optic flow resulting in linear motion can be somehow created inside the framework by alter- ing the path and velocity given in the "Shooting" scene, which includes passive locomotion. Hence, we preferred to create a separate environment for rotational optic flow which is included in several other experiments in case future researchers want to include an individual susceptibility test as a part of their experiment. [25, 26]

**Individual susceptibility test II** The Rod and Frame Test (RFT) is a perceptual test used to assess an individual's ability to judge the ver- tical orientation of an object in the presence of a tilted frame[27]. The test involves presenting a visual stimulus (usually a rod) inside a tilted frame, and the individual is asked to adjust the rod to the perceived vertical position, regardless of the frame orientation. The RFT is designed to measure an individual's ability to use visual cues to orient themselves in space, as well as their abil- ity to suppress misleading information provided by the surrounding environment. There is limited research on the use of the RFT to assess an indi- vidual's susceptibility to cybersickness. However, the RFT has been used in studies investigating the relationship between visual perception and motion sickness, which is a similar condition to cyber- sickness. Research has shown that individuals who are more susceptible to motion sickness tend to exhibit greater variability in their visual percep- tion of the vertical, as measured by the RFT[28]. This suggests that an individual's ability to main- tain a stable perception of vertical orientation may be related to their susceptibility to motion sickness. In order to support future research efforts aimed at investigating the potential of using RFT as an indicator of susceptibility to cybersickness, we have integrated this test within our framework. **Rod frame test implementation** We developed a scene for the RFT in unity3D following the reference in [28].

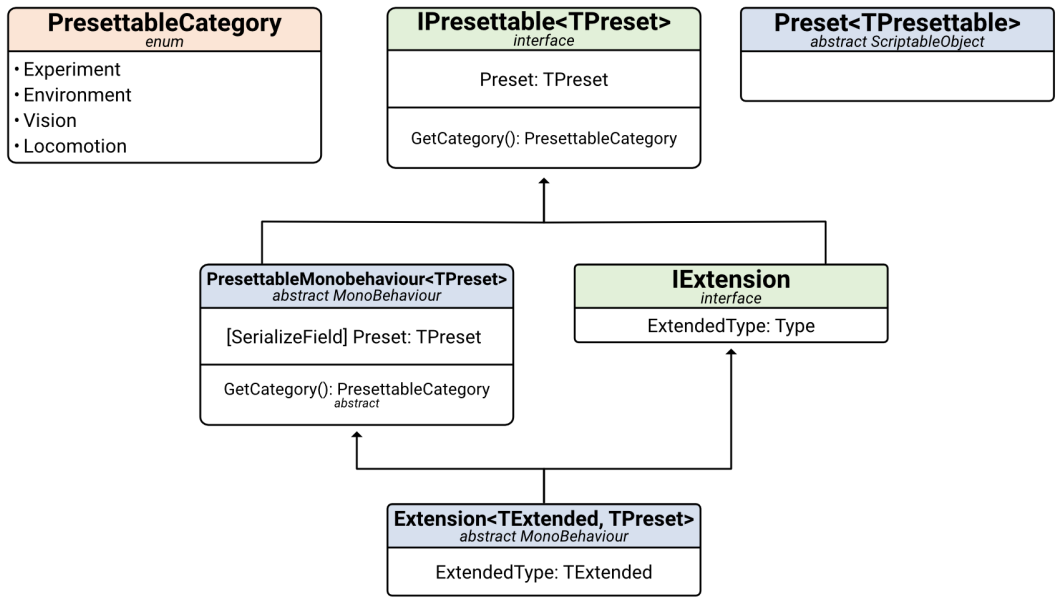

**Fig. 3**: CSAF 2.0 class hierarchy diagram

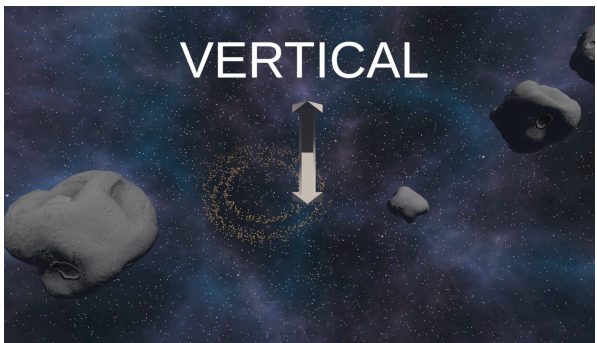

**Fig. 4**: A screenshot from sensitivity test

- A frame tilted at 18 degrees and a rod initially tilted at 27 degrees.
- With two right and left rod tilts and two right and left frame tilts, four permutations are generated. The sixteen trials are organized in a random manner.
- The rod could be rotated using the joystick. The B button is utilized to validate the response and proceed to the next trial.
- The absolute error relative to the gravitational vertical is measured for each trial.

The participants are instructed to ensure that the red rod in the VR environment is vertical relative to their own bodies, regardless of the background. When validating, users should press button B. The original rod position was similar to Figure 6a while the target rod position would be close to Figure 6b.

### 3.1.3 Environment

The environment module generally handles the game related features. The *Path Creator* contains a script that allow the researchers to design and edit their desired path for experimental purposed. The *Collectible Placer* depends on the path creator, that place a number of collectibles

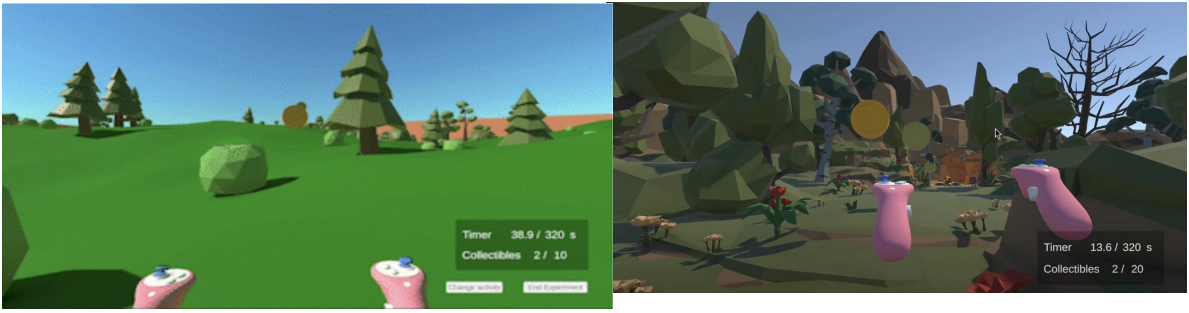

(a) Example scene with low poly forest theme and simple landscape

(b) Example scene with low poly forest theme and complex landscape

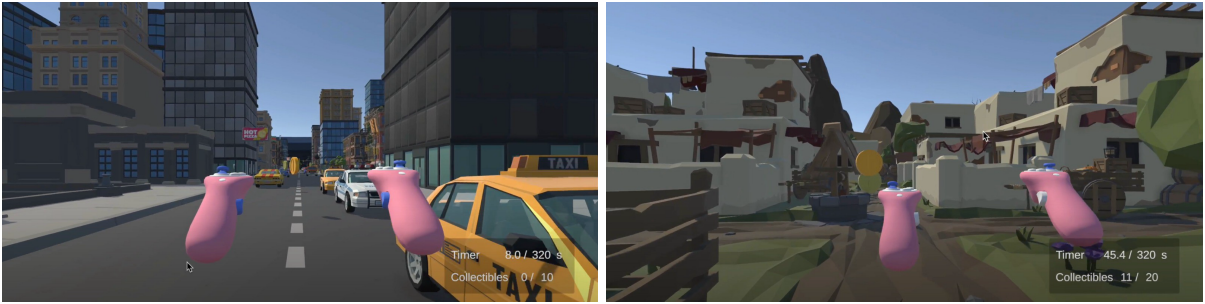

(c) Example scene with low poly city theme

(d) Example scene with low poly rural theme

**Fig. 5**: This demonstration showcases four distinct themes that can be utilized for research purposes. Additionally, it serves as a prospective database for future result comparisons and offers the flexibility for easy extension.

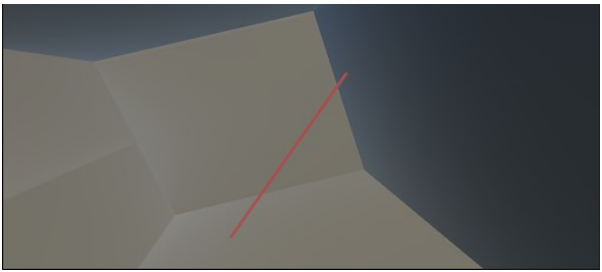

(a) Original rod position

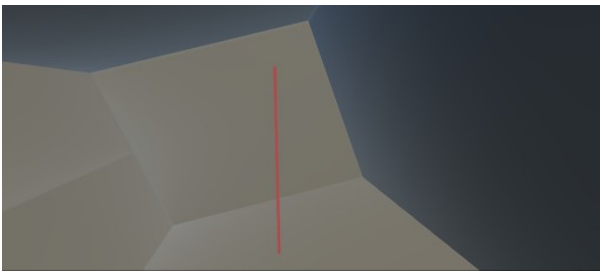

(b) Target rod position

along the designed path. Additionally, the *Terrian Modifier* provides several presets that could be modified accordingly. For example, the perlin noise is used to generate procedural terrain, where the noise values act as height maps that determine the elevation of different points on the terrain. By manipulating the parameters of the Perlin noise algorithm, such as frequency, researchers can control the overall shape and features of the generated terrain, creating natural-looking landscapes. Finally, the *Background music* offers a list of musics to serves as the ambient sound that can be looped during the gameplay. The list can be easily extended. Since background music requires looping, we provide the option of adding an initial sequence that will not get looped. This starting audio is optional. Two different copyright-free music are already included in the framework from the Game Music Starter Pack of Jac Cooper Music [29]. One of the songs is orchestral with 120 beats per minute in a calming tone. Whereas the other song is retro and upbeat with 126 beats per minute. The reason background music is added to the framework is that sound is considered as a potential influence on cybersickness [30][31],[32]. Even though experiments showed mixed results, one argument suggests that pleasing sounds for individuals could have a positive effect on cyber- sickness reduction [30, 33]. Another possibility for reduction could be the distraction of one of the senses.

### 3.1.4 Vision

The vision category includes multiple cybersick- ness reduction techniques focusing on altering the viewed content or field of view of the participant. The hypothesis and techniques included in the project are usually found in the literature.

The RestFrames option is based on the "Rest Frame Hypothesis" where keeping a stationary object in the participant's field of view no mat- ter the optic flow in the scene would prevent the sensory mismatch that results in motion sickness because the motion perceived would be seen rel- ative to the rest-frame instead of oneself [34][35] [36][37]. Some of the common objects that are used as rest-frames and could be integrated into many applications are noses, hats/helmets, and glasses. However, glasses were found to be vision-blocking and unpractical in a recent study [38], and thus the framework includes examples of a nose and some hats. However, it is possible to add a desired object in the scene and create the reference to the RestFrames preset.

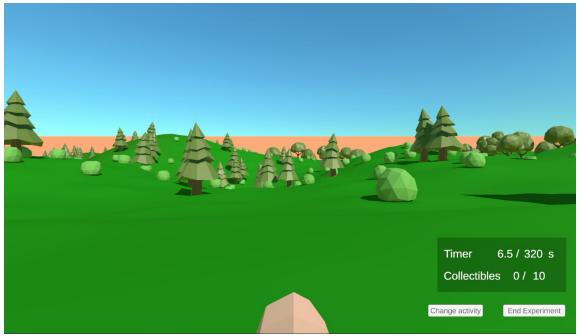

**Fig. 7**: A screenshot when nose is preset as a rest frame

The Volume components available in the scene can include other visual techniques which are: Vision Snapper, Reduced Field of View, Depth of Field, Color Manipulation, and Pixelize. Most of the mentioned techniques can be performed dynamically, except for the Pixelize effect, meaning that the effect changes depending on either the linear and/or rotational velocity of the participant inside the game.

The Reduced Field of View and the Depth of Field options are widely known tech- niques believed to be reducing cybersickness. The dynamically reduced field of view decreases the optic flow in the user's peripheral vision when the motion in the optic flow starts increasing, such as during a linear movement[39][40]. On the other hand, the latter technique blurs the vision according to the depth of objects[41][42]. Both implementations are kept unchanged from the previous version. The users are given the option to choose constant or dynamic applications of these techniques, as well as other fine tunings such as blurring intensity.

The Vision Snapper is quite an intuitive tech- nique. When a user rotates, the vision fades to black briefly as if the user had closed their eyes for a short time, which is imitating the human reflex of blinking [43][19]. This effect is better observed when the rotation is continuous.

The Color Manipulation preset is introduced to enable researchers to control a variety of effects related to colors. The intensity, contrast, and sat- uration of colors were also found to have a connec- tion to cybersickness in several studies[1][44] [45] [46]. This option only works dynamically through linear and rotational acceleration. The users can change hues for red, green, and blue colors indi- vidually or white for a more global effect. They can also modify the saturation and contrast of all colors in the scene to achieve a more general effect. There are also two features included in the project that was not ready for the VR application yet. However, the sources and scripts for the fea- tures remains to demonstrate potential extension of this framework with newly developed technolo- gies. One of these features is the Pixelize which decreases the resolution in the scene, and thus can achieve a style change from high resolution to pix- elated view. The visual realism of 3D scenes has also been an active research topic, with its impact on cybersickness being unclear[47][48]. The users can tune the pixelization effect by controlling the screen height value. Unfortunately, this effect does not work when an HMD is connected due to screen size being unreachable or incompatible.

The other feature is the Color Blur inspired from the implementation in the GingerVR project [19]. This features aims to blur the view during an active motion except for salient pixels. By blur- ing based on saliency, the technique decreases the optic flow in vision while keeping the important and attantion seeking parts untouched for the par- ticipant. Unfortunately, the saliency shader do not work with the custom post processing volume. However, it can be integrated into objects.

### 3.1.5 Locomotion

Three linear movements and two rotation alterna- tives are available for locomotion, as described in the following:

- **Continuous movement:** The player can move in any direction on the *xz* plane using the joystick of the controller.
- **Teleportation movement:** Pulling the joy- stick forward spawns a teleportation ray. When releasing, one get teleported where the ray was pointing to.
- **Grab movement:** When pressing the grip key, one can pull ourselves in any direction by moving the controller.
- **Continuous rotation:** The player can turn smoothly along the *y* axis using the joystick.
- **Snap turn rotation:** Pulling the joystick to one side makes the player turn a specified amount of degrees along the *y* axis.

In addition to these, we also created a path following the locomotion option. Given a path, the player will follow it as if they were inside a roller coaster at an adjustable speed.

## 3.2 Comparative analysis

### 3.2.1 Genericity

- **Maintainability**
  Presets are not scene or name dependent any- more. It makes them much more robust.
  Adding a category is also easier; it simply consists of creating a new enum case.
- **Stability**
  The new framework uses Unity built-in con- structs such as ScriptableObjects for its sav- ing system. This makes it more stable and bug-proof. From our tests, it can not be crashed or put in an invalid state.
- **Usability**
  The presets can now be directly visualized and edited from the master window. They can also be shared by sending the corresponding ScriptableObject file.
  They are sorted and listed by the scripts they're attached to so that one knows to what it refers to.

### 3.2.2 Trade-offs

The way the framework is constructed prevents us to have presets for entire categories or scenes. Having presets only for scripts was an acceptable trade-off in our opinion, as the ones existing before were not really usable.

To have more control over the master window, such as displaying all existing applicable presets to any GameObject, we needed to have a clean definition of which presets are available for which script so that one can list them in the window. To do so, we had to base the framework around a class hierarchy which the scripts need to inherit from.

## 3.3 Setup windows in Unity

The goal of the main setup window is to centralize all the relevant information inside a single window to facilitate the work of the users.

It should provide the general functionalities:

- Automatically detecting all GameObjects for which there exists an extension and lists them in the correct category.
- Automatically detecting all existing categories and creates a tab for each of them.
- The toggle effectively adds or removes the com- ponent from the Scene GameObjects.
- The preset is applied to the corresponding extension.

The appearance of the setup window can be found in figure 2. From the top to the bottom, the window is composed of the following elements:

1. A scrollable list of selectable panels correspond- ing to each category a presettable may belong to.
2. Two text fields and buttons to create custom features. If an invalid C# class name or an

already existing one is entered, the *Create Presettable* button is disabled. If a class that does not exist for the extended type is entered, the *Create Extension* button is disabled too.

3. All non-extensions are then displayed. Non-extensions are all scripts deriving from PresettableMonoBehaviour but not Extension. The created objects for non- extensions can be removed by deselecting the object in the setup window. The non- extensions are then ordered starting from in-scene features to not-in-scene ones.
4. At the bottom, all scripts present in the scene for which an extension exists are shown, and one can manage their extension from here as well.

Note that one can select and edit all the pre- sets from the window as well. The goal is to reduce the requirement to open the inspector window as much as possible. However, an important part of the process is performed outside the setup win- dow: creating a preset. If we had allowed the user to do so, we would not know where to place the created preset in the folder hierarchy, which would pollute the project. Instead, the user has the responsibility to create the presets as any other ScriptableObject, in the folder of their choice.

### 3.3.1 Testing Hardwares

This framework has undergone testing with two distinct virtual reality (VR) headsets and is adapt- able for use with other headsets and controllers. The study evaluated the VR experience using the HTC Vive Pro Eye headset (HTC, 2019) paired with HTC controllers. Additionally, the frame- work was tested with the Oculus Quest 2 headset (2021) along with Oculus controllers.

### 3.3.2 Technical design

To overcome the name-dependence issues men- tioned in section 2.2, a constraint was introduced for creating new custom features. When a user wants to include a new script in the framework, their script needs to follow the inheritance guide- lines provided by the framework. This helps with the stability of the framework by having control over the serialization and using Unity built-in con- structs. As a result, the setup window can list all available presets for a specific script at the cost of decreased freedom for the user.

Figure 3 visually represents the class diagram of the framework. Each class and interface is explained below in detail.

- PresettableCategory
  This enum defines the existing categories of the framework. Adding a new category only con- sists of adding a new case to it. By default, we kept the categories as follows: Experiment, Environment, Vision and Locomotion
- IPresettable<TPreset>
  This interface represents an object for which a preset exists. Its type parameter TPreset is the type of its preset. It defines a getter for the preset and the category to which it belongs.
- Preset<TPresettable>
  This abstract class is the parent of all presets in the framework. It extends Unity's built-in ScriptableObject class. ScriptableObjects are persistent data containers, easy to create from the inspector, and they perfectly fit our needs for the presets.
  Its type parameter is the IPresettable object to which this preset can be applied.
- PresettableMonoBehaviour<TPreset>
  Abstract MonoBehaviour implementation of the IPresettable interface. It provides all the clas- sic MonoBehaviour features, augmented with the serialized preset field and the abstract cat- egory to define in its children.
- IExtension
  A particular type of IPresettables. They are used to extend another script. Having such a distinction allows us to display in the setup win- dow all the available extensions for an object in the scene. For example, if one defines a CameraRotator extension that has a camera as its ExtendedType, then this extension will be proposed in the window for every camera present in the scene.
- Extension<TExtended, TPreset>
  Abstract implementation of IExtension where ExtendedType is TExtended which also extends PresettableMonoBehaviour<TPreset>

For simplicity, we did not include the IPresettable interface (without any type param- eter) in the discussion because it is only there to facilitate the use of reflection in the window. Tech- nically, this interface and the IExtension are not required, but they make reflection much easier and are not visible to the users of the framework.

## 3.4 Design of the locomotion system

### 3.4.1 Previous locomotion system

The previous LocomotionHandler only had fixed serialized fields for arbitrarily chosen LocomotionProviders which are:

- ExtendedDynamicMoveProvider, for continuous movement
- TeleportationProvider, for discrete movement
- ContinuousTurnProvider, for continuous rota- tion
- SnapTurnProvider, for discrete rotation
- TwoHandedGrabMoveProvider, for grab movement

A more in-depth explanation can be found in Viegas Milani's report [22]. The issue with this handler is that it is not expandable. If the user wanted to add their own LocomotionProvider, they would have to modify the script itself and the editor for it, which is not trivial. We had to find a way to make a more generic solution that can handle an arbitrary number of providers and offer the possibility to add or delete dynamically at runtime.

### 3.4.2 New locomotion system

To overcome this problem, we designed a new locomotion system with two List<LocomotionProvider> serialized fields; one for each controller. One can basically drag and drop any custom locomotion provider, and it should be ready to use in play mode. Again, we had to define a class hierarchy that one needs to inherit from in order for a cus- tom locomotion providers to be usable by the LocomotionHandler. One can find a visualiza- tion of this hierarchy in Figure 8. The idea is that one has to explicitly indicate what are the actions related to LocomotionProvider for each controller. Then, the LocomotionHandler can adequately enable or disable these actions based on which one is activated. It also has the advan- tage to be fully integrated into the preset system; the LocomotionHandler itself is Presettable but all the custom LocomotionProviders one defines also are.

The new LocomotionHandler works in multi- ple steps:

1. Delete all LocomotionProviders in the scene, to make sure the only ones active are those which are marked.
2. Remove from its list of providers anyone which is not deriving from ICustomLocomotionProvider. It is needed because, as our custom type is an inter- face, we cannot make it a SerializeField. So we were/are forced to pass general LocomotionProvider and then filter them.
3. Instantiate each validated provider.
4. Enable the left actions of all the providers in the list for the left controller. Do the same for the right ones.

Furthermore, the LocomotionHandler also needs the appropriate controller prefabs. As loco- motion providers require different controllers such as ray interactors, direct interactions, or no inter- actor at all. We need to pass them to serialized fields as well. To do so, we created multiple pre- fabs for any kind of interactions implemented in our project.

We also created a custom inspector for it. From the LocomotionHandler inspector, one can mod- ify the presets from all the locomotion providers attached.

## 4 Standard Workflow

The framework was designed to maximize sim- plicity and usability, catering to researchers with varying levels of proficiency in Unity coding. It accommodates both those lacking coding skills and researchers with substantial programming experience who seek to extend its functionality. In this context, we have presented an exam- ple workflow intended for experiments that do not require involvement in adding new scripts or functionalities into the framework, see Figure 9.

Researchers are advised to initiate their work- flow by making a decision regarding the appro- priate locomotion style to employ, taking into consideration the specific requirements of their experiments. This choice is between passive or active locomotion. We have created two example scenes to showcase two different styles of movement. Researchers were able to easily locate either the coin scene or the shooting scene by searching within the asset.

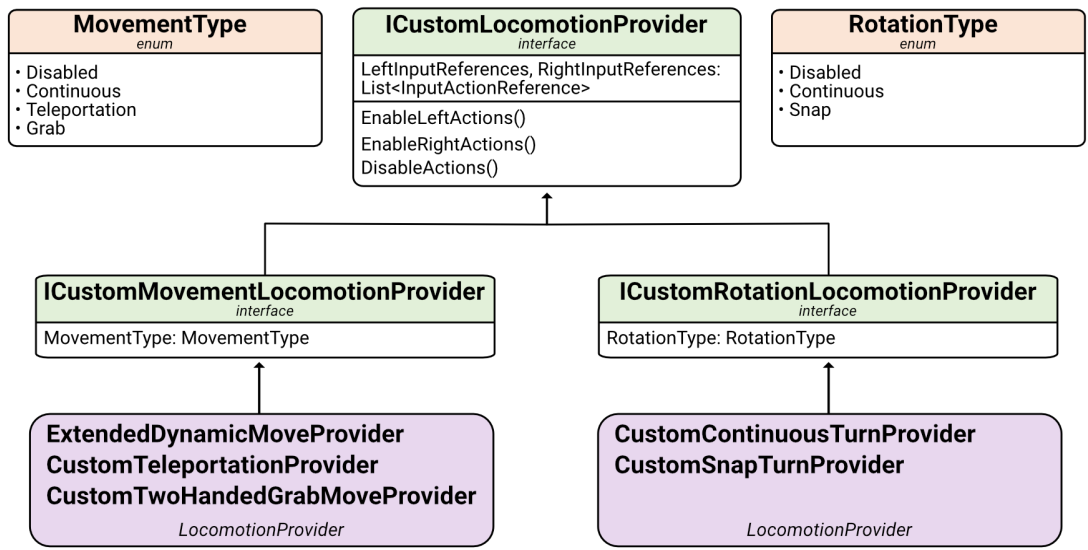

**Fig. 8**: New locomotion system class diagram

Following this decision, researchers can pro- ceed to select a scene that aligns with their desired design and complexity criteria. Subse- quently, researchers can easily enable or disable specific features within the framework, such as locomotion types, techniques aimed at reducing cybersickness. These selections should be based on the particular experimental design they have established.

Alternatively, if the existing features within the framework fail to meet the specific require- ments of the experiment, such as the need to test a new cybersickness reduction technique, researchers will have to refer to the provided guidelines for creating presettable scripts. These guidelines can be found in the ReadMe file, acces- sible through the project's Github page at the following link: https://shorturl.at/louzX. Once the newly added features have been completed, researchers will be able to utilize them as orig- inal components within the framework, further expanding the range of available options.

We have included two example videos that demonstrate the basic usage of the framework and the process of creating a script for implementing the new technique called "pixelazation" to poten- tially reduce cybersickness in the supplementary materials.

## 5 Discussion

In addition to the lack of detailed reporting in previous papers, we have also identified a con- cern related to cybersickness reduction techniques. While authors claim their effectiveness in their papers, one often faces difficulties in accessing the original designs, making it challenging to repro- duce the exact results. Furthermore, when one attempts to recreate and test these techniques with their framework, one finds that some of them lead to unnatural user experiences and could be very distractive(For example, the nose). This issue highlights the importance of transparency and accessibility in the field of cybersickness studies. Without access to the underlying implementa- tion of these techniques, researchers and develop- ers face limitations in comprehending their inner

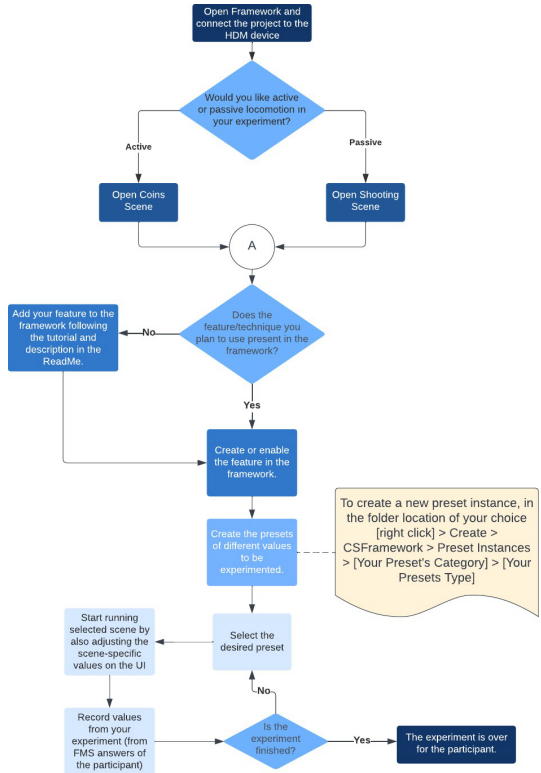

**Fig. 9**: An example workflow of the framework during an experiment

workings, adapting them to specific scenarios, or further improving upon their design.

## 5.1 Data standardization

To enhance the sharing of research findings, we have also established a standardized reference doc- ument for future studies. This document serves as a comprehensive summary, enabling researchers to easily provide detailed information about their studies. The purpose of this initiative is to facil- itate meta-analysis, a process that involves com- bining and analyzing data from multiple studies to gain broader insights. We listed in Table 2

## 5.2 Limitation and Future work

### 5.2.1 Locomotion features

Locomotion plays a pivotal role in reducing cyber-sickness, and the framework stands to benefit greatly from the incorporation of a broader range of locomotion techniques.

Video games serve as indispensable tools for exploring and refining VR locomotion tech- niques, offering valuable inspiration in the realm of cybersickness reduction. Notably, *Ready At Dawn*'s game, *Lone Echo*, has garnered posi- tive reviews for its locomotion methods, featuring player-controlled jets and zero-gravity movement through object manipulation. Introducing these locomotion techniques into our framework would be interesting, as it would facilitate actual exper- iments and enable comparisons with the existing repertoire of locomotion techniques at our dis- posal.

### 5.2.2 Improve the user interface

Currently, the user interface (UI) employed for experiment parameter control and main menu navigation are poorly designed, relying solely on default Unity components to meet functional requirements. However, the objective is to estab- lish a modular UI design that allows researchers to effortlessly integrate their own sections. By facilitating the addition of custom sections by researchers, the framework's flexibility and usabil- ity will be significantly improved. Consequently, this redesign will contribute to a more seamless and efficient workflow in conducting experiments, ultimately benefiting the general user experience and facilitating framework adoption.

### 5.2.3 Establish experiment sets

At the moment, for an experiment to be done for many participant, the researchers need to re-run the game in each conditions per participant. The preset values are not dynamically updated dur- ing runtime. Hence, for each preset values to be tried by a participant, the researcher must re-run the game. They will use already established pre- sets in each time. This might result in meaningless repetitions and annoyance over time.

An experiment set system could be created similar to a graph diagram, where the researchers draw which scene with which parameters to be played after each one. Conditions and controls in between could also be added. Such a design would automate the process of setting up for each participant in the same experiment and save the researchers from manual repetitive work.

| List of provided features | Explanation | Example |
|---|---|---|
| **Basic Demographics** | | |
| Number of participants | Total number of participants | 40 |
| Number of Female | Total number of females among participants | 20 |
| Number of experienced Users | Total number of experienced Users of VR | 10 |
| Number of inexperienced Users | Total number of inexperienced Users of VR | 8 |
| Age range | Report of age with range, mean and std | 18-36 (M: 24.5, Std: 3.45) |
| **Experiment settings** | | |
| Experiment design | The experiment design refers to the overall structure and plan for conducting a scientific experiment | Between-subject design, within-subject design, or factorial design etc. |
| Number of Sessions | Total number of sessions that each participant experienced | 2 (S1, S2) |
| Duration of baseline (If any) | Total time of each session without VR exposure | 5 min |
| Duration of the experiment (per session) | Total time of each session during VR exposure | 20 min |
| Break between sessions/exposure (If any) | Gap between each session or each period of exposure | 30 minutes |
| Break down of linear acceleration and rotation in time (If passive navigation) | Duration of time in linear acceleration and rotation | S1: Linear acceleration along longitudinal - 10 min, Rotation along Roll axis - 2 min, Rotation along Yaw axis - 6 min |
| VR content | VR Game or 360 video with names if any | Customized VR game |
| Control type | Passive locomotion or Active locomotion with controllers | Passive locomotion |
| Navigation type (per session) | Navigation type in each session | S1. Teleportation, S2. Standard control |
| Optic Flow magnitude (per session if available) | If available, provide the mean or median of optic flow magnitude | |
| **Cybersickness reduction techniques (if any)** | | |
| Name of the Techniques | Describe the technique in short | FOV reduction |
| Apply condition | When will the techniques be applied | Active when linear acceleration/deceleration |
| Details of the techniques | FOV reduction size (minimum 60 degrees) and speed (0.2 degrees/s) | |
| **Hardware settings** | | |
| HMD device | | |
| Related FOV | | |

**Table 2**: The standardized data report

### 5.2.4 Additional Cybersickness reduction techniques

Moreover, one has the option to include addi- tional cybersickness reduction techniques, such as Reversed Optic flow[49]. Nevertheless, we are cautious about the potential visual clutter or dis- traction this effect might introduce. Therefore, ensuring its natural integration becomes a cru- cial concern. One potential solution could involve transforming the reversed lines into objects within the scene that align with the overall aesthet- ics. For instance, in an ocean-themed game, the reversed lines could be replaced by fish school models, while in an open-world setting, they could be represented as birds or small flying creatures, seamlessly blending with the game environment.

## 6 Conclusion

We have proposed the CS Assessment Frame- work 2.0, whose goal is to assist cybersickness research by providing core utils and a common environment setup for experiments. It contains several locomotion methods, techniques for reduc- ing visually caused cybersickness, environmental customization, and, most importantly a frame- work that can be customized through a centralized interface. This unique control point improves the researcher's experience by allowing them to focus on the experiment parameters rather than the technical parts.

We provide an assortment of visual techniques aimed at reducing cybersickness, thereby provid- ing researchers with a greater diversity of experi- mental alternatives. Additionally, the integration of additional locomotion techniques becomes pos- sible in future iterations, as they can be easily incorporated. Furthermore, a new experiment fea- turing passive locomotion was introduced, further enriching the framework's capabilities and exper- imental scope. Moreover, a personal susceptibility test and its unique scene are available.

Finally, the proposed data standardiza- tion should facilitate the open-source research beyond the present framework. By sharing well- documented and open-source implementations, researchers can validate and build upon each other's work, fostering a collective effort to develop more practical and effective techniques. Therefore, moving forward, we emphasize the importance of encouraging authors to provide comprehensive details and open-source implemen- tations of their cybersickness research. By doing so, we can collectively work towards overcoming the challenges posed by cybersickness.

# 7 Appendix A: Glossary

- **Assets**: In Unity, assets are resources such as 3D models, textures, audio files, scripts, and other media elements used in the game development process to create the content and functionality of a game or application. In each application, the "Assets" folder serves as the root of the project's asset hierarchy.
- **Game handler**: In Unity, a game handler typically refers to a script or component that manages and controls the overall flow of the game, handling tasks such as game start, level progression, player lives, scoring, game over conditions, and other essential game mechanics.
- **GameObject**: In Unity, a GameObject is a fundamental building block that represents any element in the game world, such as characters, objects, or lights, and can have various components attached to define its behavior and appearance.
- **Manager**: In Unity, a Manager typically refers to a script or component that handles specific global functionalities or systems, such as managing game states, input, audio, or other aspects of the game to keep code organized and maintain efficient control over the game's behavior.
- **Monobehavior**: "MonoBehaviour" is a base class in Unity that enables scripts to define the behavior and functionality of game objects, making them interactive elements within the game world.
- **Presets**: In Unity, presets refer to pre-configured settings or values that can be applied to game objects or components to quickly and easily change their properties or behavior.
- **Serialization**: In Unity, Serialization refers to the process of converting variables or data within scripts into a format that can be stored and inspected in the Unity Editor, allowing developers to easily modify and save those values without altering the script's code directly.
- **Script**: In Unity, a script is a set of instructions written in a programming language (such as C#) that defines the behavior and functionality of game objects and elements within a game or application.
- **Unity**: Unity is a powerful and popular cross-platform game development engine and integrated development environment (IDE) used to create interactive 2D and 3D experiences for various platforms.

Project info: project's Github page at the following link:

https://shorturl.at/louzX Shooting scene demo I

https://youtu.be/ilCorUC3MbU Coin scene demo I:

https://youtu.be/zTCuf0tR3i0 Coin scene demo II:

https://youtu.be/64LJga0ryZo CSAF tutorial I: without coding

https://youtu.be/tItwXjdp9Wk CSAF tutorial II: with coding

https://youtu.be/wDMlGK2fE88

## Declarations

- Funding: The present research is funded by the Swiss National Science Foundation (SNF) Sinergia grant CR-SII5 180319.
- Conflict of interest: All authors have no conflict of interest to declare.
- Ethics approval: Not applicable. This paper does not involve human or animal studies.
- Consent to participate: Not applicable. Reason as Ethics approval
- Consent for publication: Not applicable. Reason as Ethics approval
- Availability of data and materials: All the video materials submitted along with this paper are available for open access.
- Code availability: Code is uploaded to Github for open access.
- Authors' contributions: Nana Tian was the initiator of this framework and made contributions to the whole manuscript text. Elif Kurtay and Dylan Vairoli contributed to multiple sections and played a crucial role in leading the implementation of the latest framework. Adriano Viegas Milani and Nana Tian worked together on the initial version of the framework and provided valuable insights for the latest revision. Additionally, Adriano Viegas Milani and Ronan Boulic conducted a thorough review of the manuscript.